\documentclass[aps,prl,twocolumn,superscriptaddress,nofootinbib]{revtex4-2}
\usepackage{graphicx}
\usepackage{amsmath,amssymb}
\usepackage{hyperref}
\usepackage{color}
\usepackage{physics}
\usepackage{float}
\usepackage{microtype}
\begin{document}

\title{Sensitivity of Electromagnetic Moments to Nuclear Forces}

\author{Jose M. Munoz}
\author{Antoine Belley}
\affiliation{Massachusetts Institute of Technology, Cambridge, Massachusetts 02139, USA}
\author{Andreas Ekstr\"om}
\affiliation{Department of Physics, Chalmers University of Technology, SE-412 96 G\"oteborg, Sweden}
\author{Gaute Hagen}
\affiliation{Physics Division, Oak Ridge National Laboratory, Oak Ridge, Tennessee 37831, USA}
\affiliation{Department of Physics and Astronomy, University of Tennessee, Knoxville, Tennessee 37996, USA}
\author{Jason D. Holt}
\affiliation{TRIUMF, Vancouver, BC V6T 2A3, Canada}%
\affiliation{Department of Physics, McGill University, Montr\'eal, QC H3A 2T8, Canada}%
\author{Ronald F. Garcia Ruiz}
\affiliation{Massachusetts Institute of Technology, Cambridge, Massachusetts 02139, USA}
\date{\today}

\begin{abstract}
Nuclear electromagnetic moments can be precisely measured across the nuclear chart, yet understanding how they emerge from the underlying nuclear force is an open challenge for nuclear theory. Using a global emulator of ab initio calculations, we show that, unlike binding energies and charge radii, 
magnetic and quadrupole moments probe components of the chiral interaction that vary strongly with neutron number. We demonstrate this sensitivity with the properties of calcium isotopes, where measured moments reveal information that is largely hidden from bulk observables. We then demonstrate a method to address the inverse problem, using selected experimental magnetic moments to constrain and reorganize the allowed parameter space of the nuclear force, improving predictions beyond the calibration set while preserving descriptions of other nuclear properties.
\end{abstract}

\maketitle

\textit{Introduction.} Nuclear electromagnetic moments and masses are among the most precisely measured properties of atomic nuclei and provide stringent, complementary probes of nuclear structure~\cite{Neyens2003,Yang2023,Dob25}. Magnetic dipole moments ($\mu$) encode the spin and orbital motion of protons and neutrons, whereas electric quadrupole moments ($Q$) are sensitive to the deformation of the nuclear charge distribution and the onset of collectivity \cite{Caprio22,Dob25}. Although recent developments in nuclear theory have allowed the reproduction of these
observables with increasing accuracy \cite{miyagi2024,Sassarini_2022,Wang2026,Ver22,Dob25,Kar24}, it remains a challenge to understand how these properties are connected to specific components of the underlying nuclear force.

For decades, comparing these observables with theory required adjustable
effective $g$-factors and effective proton and neutron charges, introduced to compensate for missing correlations and currents in truncated calculations~\cite{CastelTowner1990,garcia2015ground,Neyens2003,Dob25}. Microscopic calculations have tackled this problem by including many-body currents into the description of nuclear properties \cite{Pas13}. Nuclear interactions and electromagnetic currents can be derived consistently within Chiral Effective Field Theory ($\chi$EFT)~\cite{RevModPhys.81.1773,Machleidt:2011zz,Hammer:2019poc}, including two-nucleon currents that are essential for magnetic observables. These advances have enabled calculations of magnetic dipole moments from light to medium-mass nuclei~\cite{Martin2023,miyagi2024,acharya2024magnetic,4tky-t2h1},
similar to the resolution of the long-standing $g_A$-quenching puzzle in beta decay~\cite{Gysbers2019}.

\begin{figure*}[t!]
    \centering
    \includegraphics[width=1.0\linewidth]{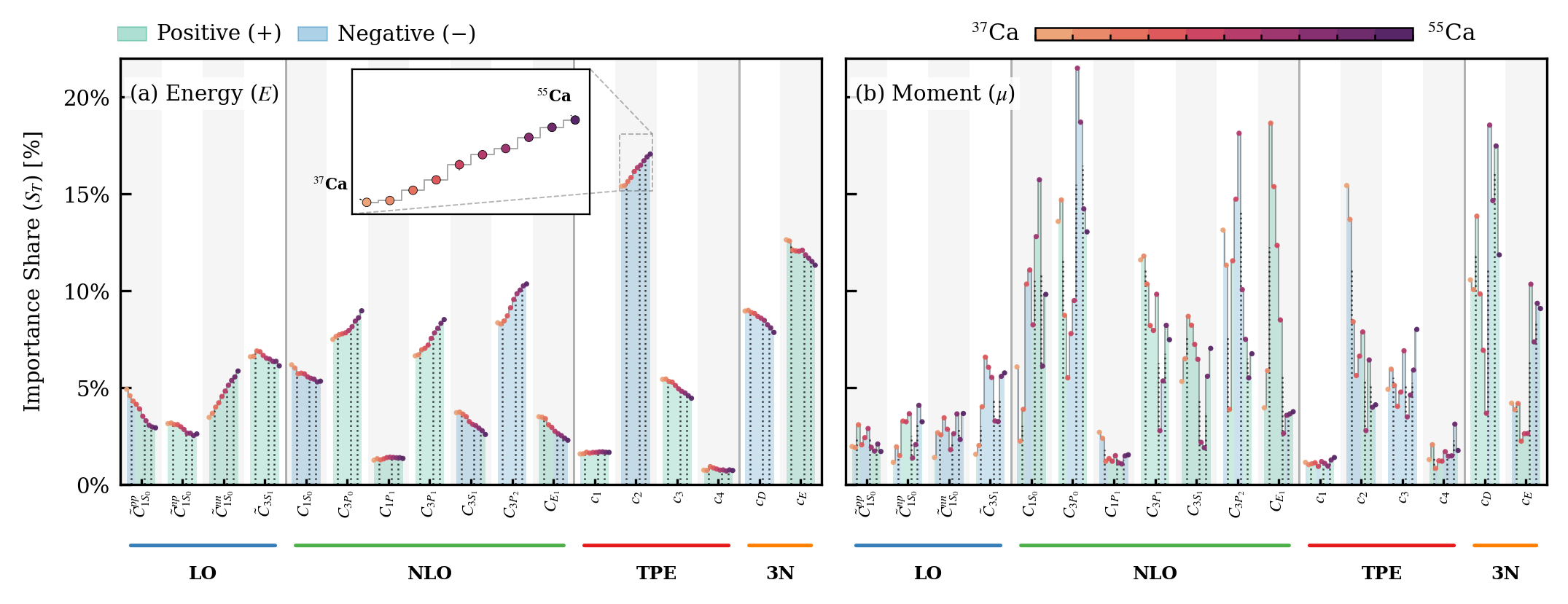}
    \caption{\textbf{Contrasting information content of bulk and electromagnetic observables.}
    Posterior-integrated importance share of each LEC for the binding energy ($E$) and magnetic dipole moment ($\mu$) at each odd-$A$ isotope $^{37\text{--}55}$Ca.  Green and blue bars indicate average positive and negative LEC contributions to the
observable value respectively.
    For the bulk observables (left) the relative change in importance ($|\Delta \text{share}|$) across isotopes is about an order of magnitude smaller than for the EM moments shown on the right panels. 
    For the magnetic moments (right) the pattern restructures from isotope to isotope, with an average change more than an order of magnitude larger. 
    }
    \vspace{-10pt}
    \label{fig:SharedScales}
\end{figure*}

Reproducing data, however, is not the same as understanding their origin. A key open challenge is to identify which components of the nuclear force control the magnitude of electromagnetic moments, their evolution along isotopic chains, and their abrupt changes near shell closures. These are the central questions that we address in this work.

In $\chi$EFT, nuclear forces are parameterized by Low-Energy Constants (LECs), calibrated to nucleon-nucleon scattering and few-body observables~\cite{PhysRevC.91.051301,carlsson2016uncertainty,jiang2024emulating}. Establishing a microscopic connection between nucleon-nucleon interactions and nuclear properties therefore requires determining how nuclear observables respond to variations in the LECs over the full range allowed by existing data.  Such a mapping is computationally prohibitive with direct many-body calculations, as they demand
millions of model evaluations~\cite{ekstrom2019global,konig2020,hu2022ab},
far beyond the reach of direct state-of-the-art many-body
calculations~\cite{Hergert:2020bxy}. Here we overcome these challenges through two
main developments. First, we introduce a global physics-driven emulator, the
Fidelity-Resolved Affine Matrix Emulator (FRAME), trained on valence-space
in-medium similarity renormalization group
(VS-IMSRG)~\cite{Stro17ENO,Stroberg_2019} calculations with electromagnetic
operators transformed consistently with the Hamiltonian~\cite{Parz17Trans}.
FRAME extends parametric matrix models~\cite{Cook:2024toj} across many isotopes and model-space sizes~\cite{belley2025globalframeworkemulationnuclear},
providing fast and accurate predictions of energies, radii, and
electromagnetic moments. Second, we develop a correlated sensitivity analysis that quantifies the contribution of each LEC to each observable over the posterior ensemble of interactions compatible with existing data. Unlike conventional variance-based approaches, this analysis includes correlations among LECs, thereby identifying the components of the nuclear force that govern nuclear electromagnetic properties.

Our developments also make it possible to tackle the inverse problem. Because electromagnetic moments are measured with a precision far exceeding that of present ab initio calculations, a natural question arises: can precision
measurements of electromagnetic moments be used to constrain the LECs that
describe the nuclear force?
We use our framework to map how observables in the calcium chain
$^{37\text{--}55}$Ca respond to variations of the $\chi$EFT low-energy
constants. We find that the LEC sensitivities of binding energies and charge
radii are nearly independent of neutron number, indicating that these bulk
observables carry largely redundant information about the interaction. In
contrast, magnetic and quadrupole moments exhibit strongly isotope-dependent
sensitivities, providing nonredundant information about the nuclear force.
We then show that using the experimentally known magnetic moments of selected
calcium isotopes as constraints significantly reduces the uncertainties of
the LECs while preserving or improving the description of other
observables.

\textit{Nuclear emulator.} To rigorously explore the dependence of nuclear observables on the chiral interaction, we
developed the Fidelity-Resolved Affine Matrix Emulator (FRAME). FRAME defines a
global surrogate map $ g:(Z,N,f,\boldsymbol{\alpha})
  \mapsto
  \{E,\langle O_q\rangle\},$
where $(Z,N)$ specifies the nucleus, $f$ denotes the model-space fidelity
identified here with the VS-IMSRG single-particle truncation $e_{\max}$, and
$\boldsymbol{\alpha}$ is the vector of chiral EFT low-energy constants (LECs).
The emulator returns a consistent set of level energies $E$, and expectation values $\langle O_q\rangle$ for the charge radius, magnetic dipole moment, and
electric quadrupole moment,
$\langle O_q\rangle\in\{R_{\rm ch},\mu,Q\}$. FRAME is trained on VS-IMSRG calculations in which electromagnetic operators are evolved consistently with the Hamiltonian for the $\Delta$NNLO$_{\rm{{GO}}}$(394) interaction~\cite{jiang2020}, built from a $\Delta$-full $\chi$EFT expansion at N$^2$LO~\cite{Krebs:2007rh} with a 394 MeV regulator cutoff. It therefore provides rapid predictions across nuclei, model spaces, and interactions, while inheriting the controlled approximations and systematic limitations of the underlying VS-IMSRG calculations, including
many-body and operator truncation effects.

The LEC domain explored in this work is inherited from the chiral EFT calibrations of Refs.~\cite{belley2024ab,jiang2024emulating}. Each interaction
is specified by a 17-dimensional vector $\boldsymbol{\alpha}$. The ensemble was
constructed through {history matching}~\cite{Vernon:2010,Vernon:2014,hu2022ab}:
emulators for few-body nuclei and infinite nuclear matter were confronted with
selected calibration data, and regions of parameter space unable to reproduce
those data within a combined experimental and theoretical uncertainty budget
were iteratively ruled out. This procedure yields a cloud of $\mathcal{O}(10^4)$ non-implausible interactions that reproduce the ground-state
energies and point-proton radii of light nuclei with $A=2$--$4$. We use this cloud as the prior distribution for the LECs.

From this prior, we construct posterior ensembles by assigning each interaction
a likelihood weight according to its agreement with calibration data. We
consider two likelihoods. The first, $\mathcal{L}_{2-4}$, includes the
ground-state energies and point-proton radii of nuclei with $A=2$--$4$. The
second, $\mathcal{L}_{(2-4)+^{16}{\rm O}}$, augments this set with the
ground-state energy and point-proton radius of $^{16}$O. For a calibration
observable $y_k$, the likelihood is taken to be Gaussian,
\begin{equation}
  \mathcal{L}(\boldsymbol{\alpha})
  \propto
  \exp\left[
  -\frac{1}{2}
  \sum_k
  \frac{
  \left(y_k^{\rm th}(\boldsymbol{\alpha})-y_k^{\rm exp}\right)^2
  }{
  \sigma_{k,\rm exp}^2+
  \sigma_{k,\rm emu}^2+
  \sigma_{k,\chi{\rm EFT}}^2+
  \sigma_{k,\rm MB}^2
  }
  \right],
\end{equation}
where the total variance includes experimental, emulator, chiral EFT
truncation, and many-body truncation uncertainties in quadrature; details are
given in the End Matter. Weighted samples represent the posterior
$\{\boldsymbol{\alpha}_i,w_i\}$, with $w_i=\mathcal{L}(\boldsymbol{\alpha}_i)$
for either $\mathcal{L}_{2-4}$ or
$\mathcal{L}_{(2-4)+^{16}{\rm O}}$.

We validated FRAME against independent VS-IMSRG calculations excluded from the
training set. The held-out test set consists of non-implausible interactions at
the highest fidelity used in training, $e_{\max}=10$. The emulator reproduces
the VS-IMSRG results for all observables considered, with no visible systematic
bias as a function of mass number $A$. Validation details are provided in the
Supplemental Material. This validation is crucial because the subsequent
sensitivity and Bayesian analyses require millions of evaluations over the LEC
manifold, which would be infeasible with direct VS-IMSRG calculations.

For an observable
$Y\in\{E,R_{\rm ch},\mu,Q\}$, the posterior predictive distribution is obtained
by marginalizing the FRAME prediction over the weighted LEC ensemble,
\begin{equation}
  p(Y|{\cal D})
  =
  \sum_i \bar w_i\,
  p_{\rm emu}\!\left[
  Y\,\middle|\,
  Z,N,f,\boldsymbol{\alpha}_i
  \right],
  \qquad
  \bar w_i=\frac{w_i}{\sum_j w_j},
\end{equation}
where ${\cal D}$ denotes the calibration data and $p_{\rm emu}$ accounts for
the emulator uncertainty. In this work, posterior predictives for
energies, charge radii, magnetic moments, and quadrupole moments are shown as
bands representing the combined LEC and emulator degrees of belief for each
isotope and observable.

\textit{Information content of nuclear observables.}
Using FRAME, we quantify how the low-energy constants (LECs) of the chiral
interaction affect nuclear observables ($E$, $R_{\mathrm{ch}}$, $\mu$, and $Q$) along the Ca ($Z=20$) isotopic chain. With a proton magic number $Z = 20$ and several neutron shell and subshell closures at $N = 20, 28, 32$, and $34$, the calcium isotopic chain provides a testing ground for nuclear theory. Near these closures, many-body correlations are reduced, minimizing the dependence on the many-body method and enabling a reliable connection to the nuclear force.

To quantify the importance of the different LECs in
explaining the values of nuclear observables, we perform
a posterior-integrated local importance analysis using SHAP values~\cite{Owen}. For each nucleus $(Z,N)$, model-space fidelity $e_{\max}$, and observable
$Y\in\{E,R_{\rm ch},\mu,Q\}$, FRAME provides a fast map
$\boldsymbol{\alpha}\mapsto Y$ over the history-matched LEC ensemble. We
evaluate this map using the posterior weights obtained from
$\mathcal{L}_{(2-4)+^{16}{\rm O}}$, and quantify the contribution of each LEC
from the local importance value. For a given sample, the
Shapley value of LEC $\alpha_j$ is the average marginal change in the
prediction caused by adding $\alpha_j$, averaged over all possible orderings of
the LECs~\cite{lundberg2017unified}. We then average these attributions over
the posterior ensemble using the normalized weights $\bar w_i$. This procedure
yields a normalized importance share for each LEC, measuring the typical
magnitude of the shift in the observable driven by that coupling within the
physically supported region of parameter space.

This posterior-integrated attribution provides several advantages with respect to conventional Sobol-like variance decompositions used previously~\cite{belley2024ab,ekstrom2019global,belley2025globalframeworkemulationnuclear}.
Sobol indices are usually defined for independently distributed parameters,
often over a hypercube. Such domains include large regions of LEC space
that are already ruled out by few-body and nuclear-matter constraints and
therefore neglect the correlations induced by calibration. In contrast, our
Shapley analysis is performed directly on the posterior ensemble and therefore
respects both the correlations among LECs and the nontrivial geometry of the
history-matched manifold. 

The resulting attribution maps reveal a clear separation between bulk and
electromagnetic observables. Figure~\ref{fig:SharedScales} shows the
posterior-weighted LEC importance for $E$ and $\mu$ across
$^{37\text{--}55}$Ca, with $R_{ch}$ and $Q$ given in the Supplemental Material. For binding energies and charge radii
(left panel of Fig.~\ref{fig:SharedScales}), the LEC importance pattern shows relatively small changes with neutron number, with  $|\Delta \text{share}| < 0.4\%$. These observables are dominated by a set of couplings, primarily the two-pion-exchange coefficient $c_2$ and the 3N
couplings $c_D$ and $c_E$. Thus, along the calcium chain, bulk observables
carry largely redundant information about the nuclear interaction. This
behavior is consistent with earlier variance-based studies in oxygen,
germanium, and nickel isotopes~\cite{ekstrom2019global,belley2024ab,belley2025globalframeworkemulationnuclear}.

Electromagnetic moments display a very different pattern. The LEC
importance shares for the magnetic moments
(right panel of Fig.~\ref{fig:SharedScales}) vary strongly with neutron number and are
not controlled by a single persistent set of LECs. Instead, the attributions
fragment into several isotope-dependent structures, indicating that
electromagnetic moments probe different combinations of interaction terms as
the valence-neutron configuration changes. The average change of the
importance share between neighboring isotopes (bars in
Fig.~\ref{fig:SharedScales}) is more than an order of magnitude larger for the
moments than for the bulk observables. 

Several features of this behavior can be understood from a shell-model
perspective. The triplet-$P$ contact couplings
($C_{^3P_0}, C_{^3P_1}, C_{^3P_2}$) show a strong neutron-number dependence in
the magnetic moments, a dependence that is absent in $E$ and $R_{\rm ch}$.
Their importance changes as valence neutrons fill the $f_{7/2}$
orbital in $^{41\text{--}48}$Ca and then move into the $p_{3/2}$ and
$p_{1/2}$ orbitals in $^{49\text{--}52}$Ca and
$^{53\text{--}55}$Ca, respectively. Since magnetic moments are directly
sensitive to the spin and orbital angular-momentum content of the valence
nucleons, they amplify changes in the spin-isospin and spin-orbit components
of the interaction that are largely averaged out in bulk observables.

\begin{figure}[!h]
    \centering
    \includegraphics[width=\linewidth]{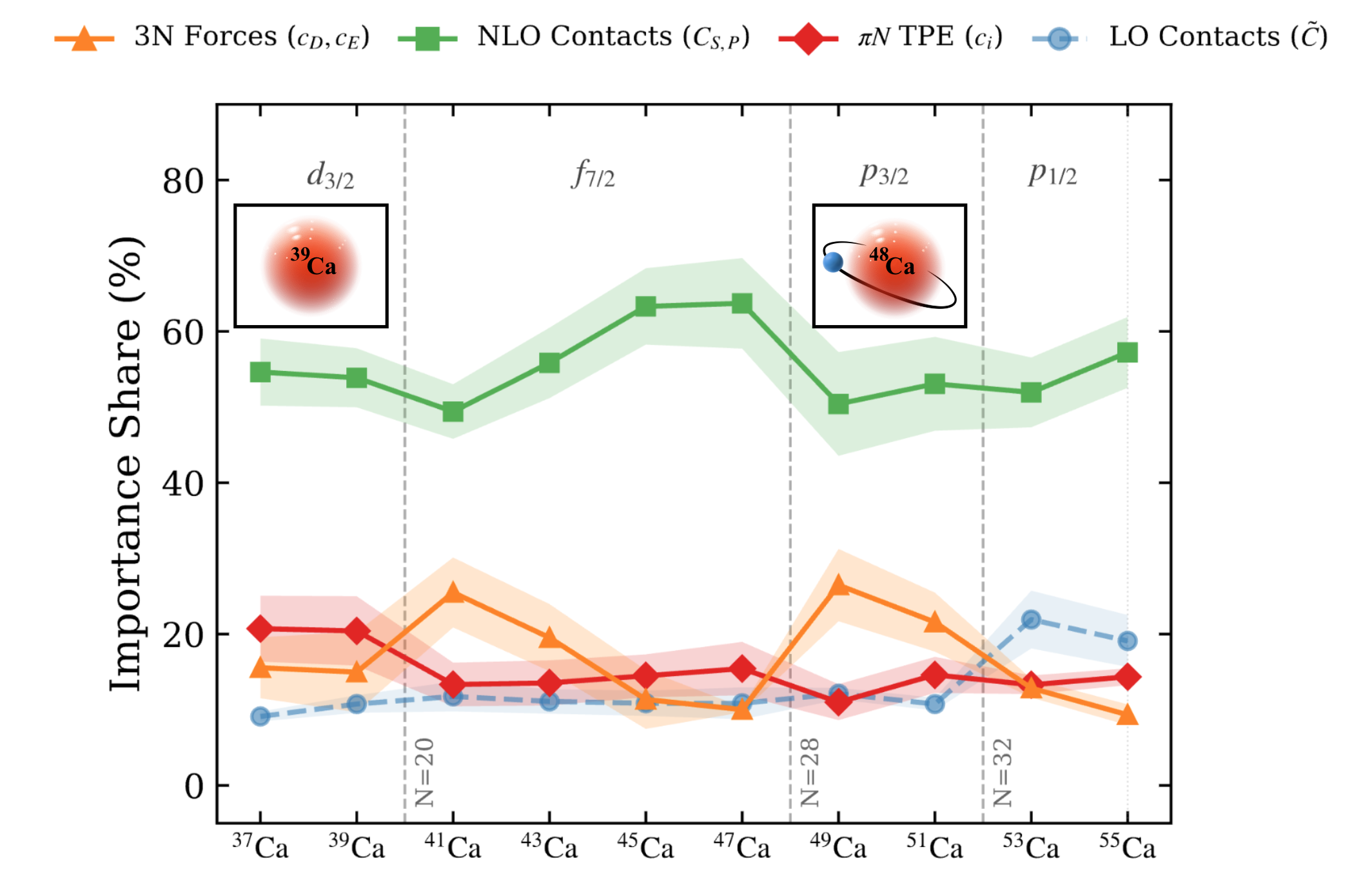}
    \caption{\textbf{Sector resolved importance for $\mu$}. Posterior-integrated importance by physical sector for $\mu$ across the calcium chain. The shaded regions represent the standard deviation within each physical sector. Representations of a nearly single-neutron-hole and single-particle neutron limit ($^{39}$Ca and $^{49}$Ca respectively) are shown.}
    \label{fig:grouped_mu_evol}
    
\end{figure}

We group individual LEC
attributions into leading and next-to-leading contact, two-pion-exchange (TPE), and three-nucleon (3N)
sectors, and trace their relative importance along the isotopic chain. Fig.~\ref{fig:grouped_mu_evol}
shows this sector-resolved importance for the magnetic moments.
As seen in the figure, the dominant contributions arise from the NLO spin-orbit contact sector, which controls
single-particle level splittings and therefore the spin and orbital admixtures
of the valence nucleons. The 3N couplings $(c_D,c_E)$  also exhibit an isotope-dependent pattern which is enhanced at shell closures, with peaks at $N=20$, $28$,
and $32$. This suggests that magnetic moments near closed shells can provide sensitive, complementary probes of three-nucleon physics.

\textit{Constraining the nuclear force with magnetic moments.} 
To study how much additional information magnetic moments provide, we perform a targeted Bayesian update of the history-matched LEC ensemble constraining to the experimental data on magnetic moments of $^{39,47,49,51}\text{Ca}$. This selection minimizes many-body truncation errors by focusing on single-particle and single-hole configurations relative to the closures at $N=20,28,32$~\cite{miyagi2024, acharya2024magnetic}. Note that our goal is not merely to construct a new global fit of the LEC values. Rather, we aim to investigate how the information identified through the importance analysis translates into a narrower and reorganized posterior distribution for the LECs. In particular, to assess how strongly the experimental magnetic-moment data constrain the LECs, how these constraints affect the uncertainties of predictions for both the magnetic moments included in the fit and those excluded from it, and how the resulting changes influence predictions for other observables.

\begin{figure}[htbp!]
    \centering
    \includegraphics[width=1\linewidth]{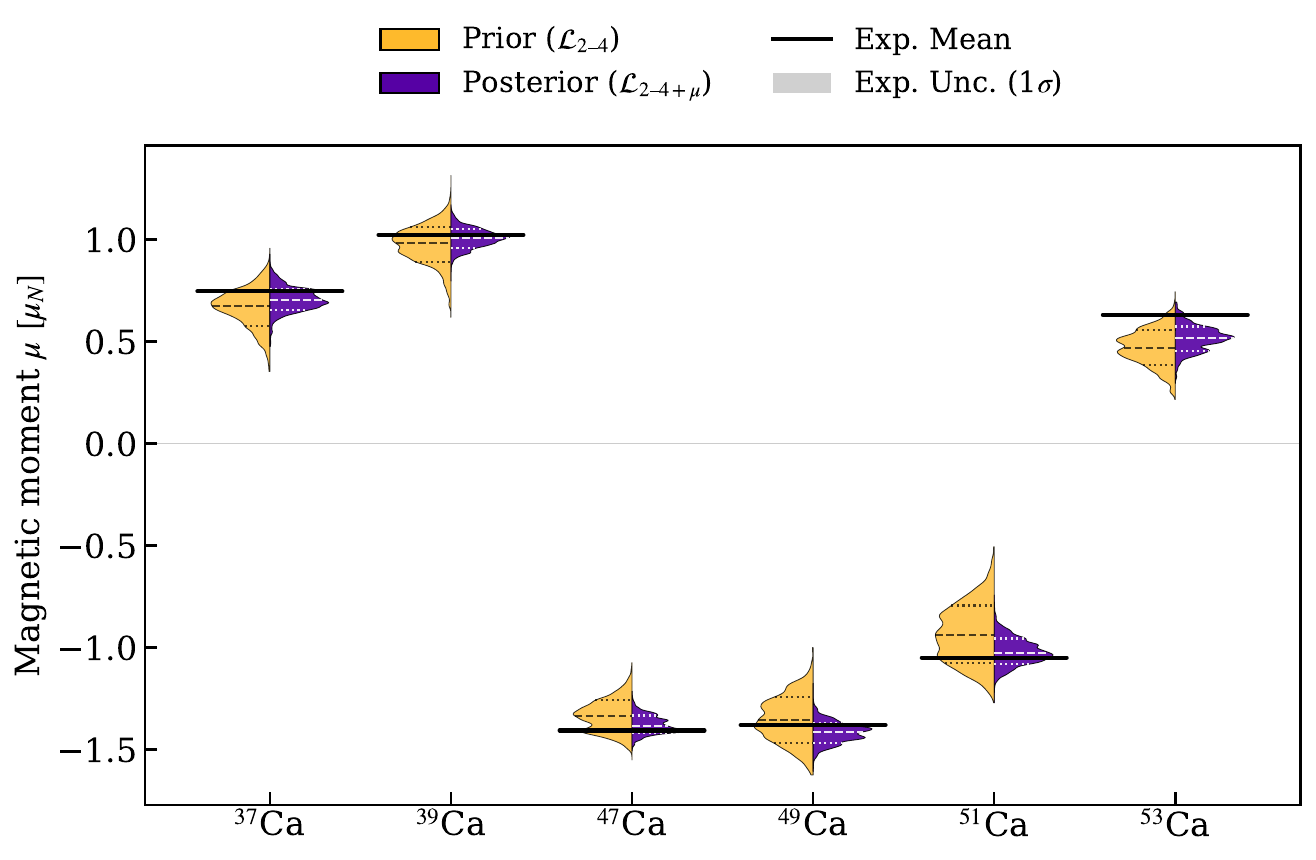}
    \caption{\textbf{Predictive posteriors for $\mu$ before and after the likelihood update}. Nuclear magnetic dipole moments for calcium isotopes in the vicinity of the neutron closed shells. 
    We compare the prior (yellow) and updated posterior (purple) distributions against experimental values (black bands).}
    \label{fig:mu_selected}
\end{figure}

Figure~\ref{fig:mu_selected} compares the prior and posterior distributions of $\mu$ for calcium isotopes near the neutron shell closures. For the calibration isotopes ($^{39,47,49,51}$Ca), the posterior narrows around the experimental values as expected. 
Importantly, the non-calibration isotopes $^{37}$Ca and $^{53}$Ca also exhibit a significant reduction in spread while remaining compatible with experiment, indicating that the constraints propagate coherently across the chain through the shared LEC dependence.

For each isotope whose experimental data are used for calibration,  we perform a Bayesian update using a Gaussian likelihood for the experimental value of each $\mu$, defining the total variance by summing all sources of uncertainty in quadrature (see End Matter). This update induces a structural reorganization of the LEC posterior. Rather than merely shrinking the parameter volume, it sharpens the correlations among couplings, identifying physical dependencies that remain effectively unconstrained by $E$ and $R_{\mathrm{ch}}$ alone. In practice, the update reshapes the distribution of LECs in a way that is both visible in low-dimensional projections but also quantifiable in the full space via reduction of the variance, $\Delta\!\log|\Sigma|\!=\!\log|\Sigma_{\rm prior}|-\log|\Sigma_{\rm post}|=3.0$. Importantly, this metric is positive across all weight-tempering schemes analyzed, while maintaining a statistically robust effective sample size (see End Matter). This shows that the magnetic moment data provide important constraints without contradicting the interaction prior.

\begin{figure}
    \centering
    \includegraphics[width=\linewidth]{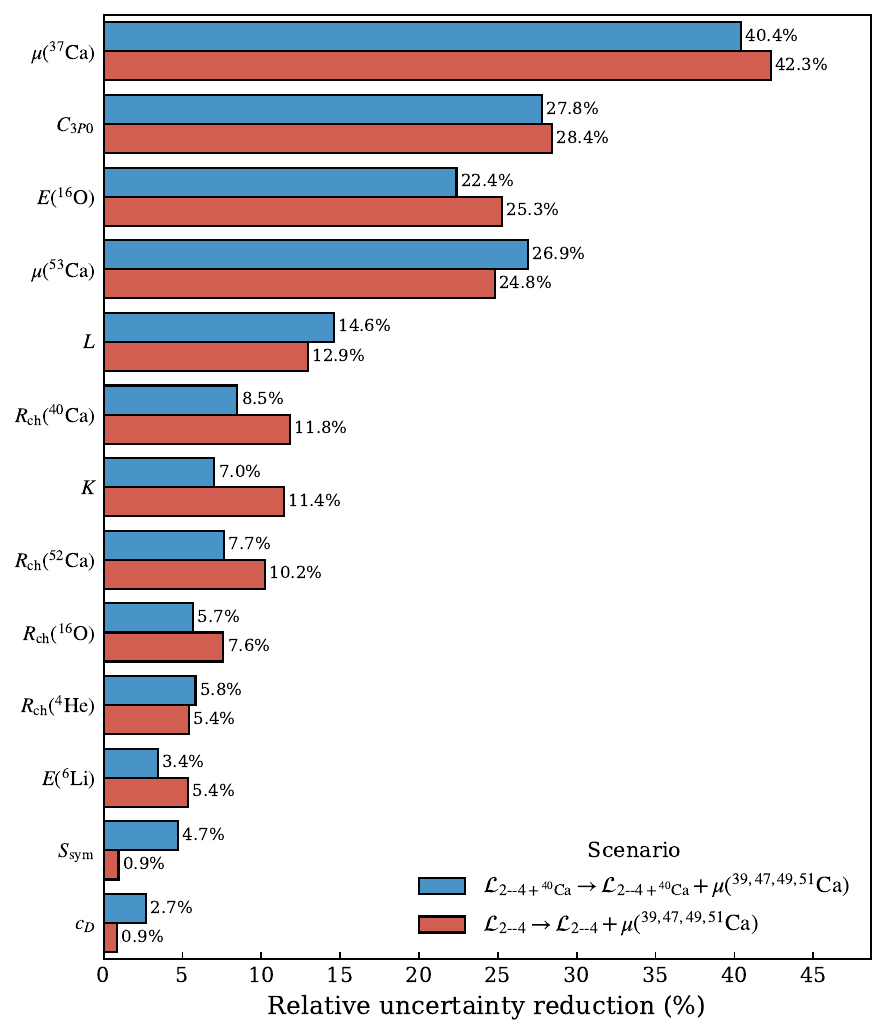}
    \caption{\textbf{Predictive band improvements after including $\mu(^{39,47,49,51}\mathrm{Ca})$}.
    Bars show the percentage reduction in posterior standard deviation,
    $1 - \sigma_{\mathrm{post}}/\sigma_{\mathrm{prior}}$,
    for observables, nuclear matter parameters (symmetry energy $S_{sym}$, its density slope L, and the incompressibility K), and selected LECs.
    The prior is the history-matched ensemble weighted by $\mathcal{L}_{2-4}$.}
    \label{fig:final_reduction_comparison}
\end{figure}

A natural question is whether this improvement remains confined to magnetic moments or extends to other observables and interaction parameters. In Fig.~\ref{fig:final_reduction_comparison} we address this by showing the reduction in the 68\% DoB predictive bands after incorporating $\mu(^{39,47,49,51}\mathrm{Ca})$, for two starting configurations that differ by the inclusion of $R_{\mathrm{ch}}(^{40}\mathrm{Ca})$ in the prior. Because the latter already tightens the calcium-relevant parameter space, improvements that persist across both configurations can be attributed to the constraint given by the magnetic moment data. The largest gains appear in the predicted moments of uncalibrated isotopes and in the NLO contact $C_{^3P_0}$, identified by the SHAP analysis as a key driver of spin-orbit splittings. Meanwhile, the predictions for bulk observables such as $E(^{16}\mathrm{O})$ and $R_{\mathrm{ch}}(^4\mathrm{He})$ remain in agreement, confirming that the magnetic moment data provide complementary constraints while preserving or improving the existing description of bulk properties.

\textit{Summary,} we presented the development of FRAME, a global emulator of ab initio VS-IMSRG
calculations with consistently evolved electromagnetic operators, enabling
rapid predictions of energies, radii, magnetic moments, and quadrupole moments
over large ensembles of chiral interactions. We combined this emulator
with a posterior-integrated sensitivity analysis that attributes observable
changes to individual low-energy constants while preserving the correlations
and geometry of the calibrated LEC manifold.

Applying this framework to describe the properties of $^{37\text{--}55}$Ca, we find marked differences in how distinct components of the nuclear force influence bulk and electromagnetic observables. The sensitivity of binding energies and charge
radii to the LECs show a smooth change with respect to the neutron number. In
contrast, magnetic and quadrupole moments display strong isotope-dependent
sensitivities that follow shell evolution and expose spin-dependent contacts
and three-nucleon contributions that are largely unconstrained by bulk observables.

These results enable us to move beyond conventional theory–experiment comparisons and formulate the extraction of information about the nuclear interaction as an inverse problem. Using selected measured calcium magnetic
moments as Bayesian constraints, we show that precision moment data sharpen
and reorganize the allowed chiral-interaction parameter space, while
preserving the description of few-body systems, oxygen, and bulk calcium
properties. Beyond providing stringent tests of nuclear many-body theory, electromagnetic moments of nuclei with simple particle–particle and particle–hole configurations offer complementary information for constraining nuclear forces. 
Thus, our developments provide a systematic framework for experimental design, enabling the prioritization of future measurements according to their expected information gain and their ability to constrain poorly understood aspects of nuclear interactions.

\begin{acknowledgments}

J.M.M. and R.F.G.R. acknowledge support from the MIT Generative AI Impact Consortium (MGAIC). This work was supported by the Office of Science U.S. Department of Energy, under grants  DESC0021176, DE-SC0021179, SciDAC-5 (NUCLEI collaboration), and DE-SC0026198 (STREAMLINE collaboration). 
The IMSRG code used in this work makes use of the Armadillo \texttt{C++} library \cite{Sanderson2016}.  Computational resources were provided by SubMIT at MIT Physics.  This research used resources from the Oak Ridge Leadership Computing Facility located at Oak Ridge National Laboratory, which is supported by the Office of Science of the U.S. Department of Energy under contract No. DE-AC05-00OR22725.
We acknowledge the support of the Natural Sciences and Engineering Research Council of Canada (NSERC) [PDF-587464-2024], and the Swedish Research Council (Grant No. 2024-04681). We thank Christian Forssén and Matthias Heinz for insightful discussions.

\end{acknowledgments}

\bibliographystyle{apsrev4-2}
\bibliography{main}

@article{Pas13,
  title = {Quantum Monte Carlo calculations of electromagnetic moments and transitions in $A\ensuremath{\le}9$ nuclei with meson-exchange currents derived from chiral effective field theory},
  author = {Pastore, S. and Pieper, Steven C. and Schiavilla, R. and Wiringa, R. B.},
  journal = {Phys. Rev. C},
  volume = {87},
  issue = {3},
  pages = {035503},
  numpages = {15},
  year = {2013},
  month = {Mar},
  publisher = {American Physical Society},
  doi = {10.1103/PhysRevC.87.035503},
  url = {https://link.aps.org/doi/10.1103/PhysRevC.87.035503}
}

@misc{Dob25,
       author = {{Dobaczewski}, J. and {Backes}, B.~C. and {de Groote}, R.~P. and {Restrepo-Giraldo}, A. and {Sun}, X. and {Wibowo}, H.},
        title = "{Electromagnetic and Exotic Moments in Nuclear DFT}",
         year = 2025,
        month = nov,
          doi = {10.48550/arXiv.2511.04632},
 primaryClass = {nucl-th},
 archivePrefix = {arXiv},
       eprint = {2511.04632},
       adsurl = {https://ui.adsabs.harvard.edu/abs/2025arXiv251104632D}
}

@ARTICLE{Ver22,
       author = {{Vernon}, A.~R. and {Garcia Ruiz}, R.~F. and {Miyagi}, T. and {Binnersley}, C.~L. and {Billowes}, J. and {Bissell}, M.~L. and {Bonnard}, J. and {Cocolios}, T.~E. and {Dobaczewski}, J. and {Farooq-Smith}, G.~J. and {Flanagan}, K.~T. and {Georgiev}, G. and {Gins}, W. and {de Groote}, R.~P. and {Heinke}, R. and {Holt}, J.~D. and {Hustings}, J. and {Koszor{\'u}s}, {\'A}. and {Leimbach}, D. and {Lynch}, K.~M. and {Neyens}, G. and {Stroberg}, S.~R. and {Wilkins}, S.~G. and {Yang}, X.~F. and {Yordanov}, D.~T.},
        title = "{Nuclear moments of indium isotopes reveal abrupt change at magic number 82}",
      journal = {\nat},
         year = 2022,
        month = jul,
       volume = {607},
       number = {7918},
        pages = {260-265},
          doi = {10.1038/s41586-022-04818-7},
       adsurl = {https://ui.adsabs.harvard.edu/abs/2022Natur.607..260V}
}

@ARTICLE{Kar24,
       author = {{Karthein}, J. and {Ricketts}, C.~M. and {Garcia Ruiz}, R.~F. and {Billowes}, J. and {Binnersley}, C.~L. and {Cocolios}, T.~E. and {Dobaczewski}, J. and {Farooq-Smith}, G.~J. and {Flanagan}, K.~T. and {Georgiev}, G. and {Gins}, W. and {de Groote}, R.~P. and {Gustafsson}, F.~P. and {Holt}, J.~D. and {Kanellakopoulos}, A. and {Koszor{\'u}s}, {\'A}. and {Leimbach}, D. and {Lynch}, K.~M. and {Miyagi}, T. and {Nazarewicz}, W. and {Neyens}, G. and {Reinhard}, P.-G. and {Sahoo}, B.~K. and {Vernon}, A.~R. and {Wilkins}, S.~G. and {Yang}, X.~F. and {Yordanov}, D.~T.},
        title = "{Electromagnetic properties of indium isotopes illuminate the doubly magic character of $^{100}$Sn}",
      journal = {Nature Physics},
         year = 2024,
        month = nov,
       volume = {20},
       number = {11},
        pages = {1719-1725},
          doi = {10.1038/s41567-024-02612-y},
archivePrefix = {arXiv},
       eprint = {2310.15093},
 primaryClass = {nucl-ex},
       adsurl = {https://ui.adsabs.harvard.edu/abs/2024NatPh..20.1719K}
}

@article{Caprio22,
  title = {Robust ab initio prediction of nuclear electric quadrupole observables by scaling to the charge radius},
  author = {Caprio, Mark A. and Fasano, Patrick J. and Maris, Pieter},
  journal = {Phys. Rev. C},
  volume = {105},
  issue = {6},
  pages = {L061302},
  numpages = {7},
  year = {2022},
  month = {Jun},
  publisher = {American Physical Society},
  doi = {10.1103/PhysRevC.105.L061302},
  url = {https://link.aps.org/doi/10.1103/PhysRevC.105.L061302}
}

@article{konig2020,
  title = {Eigenvector continuation as an efficient and accurate emulator for uncertainty quantification},
  volume = {810},
  issn = {0370-2693},
  url = {https://www.sciencedirect.com/science/article/pii/S0370269320306171},
  doi = {https://doi.org/10.1016/j.physletb.2020.135814},
  journal = {Phys. Lett. B},
  author = {König, S. and Ekström, A. and Hebeler, K. and Lee, D. and Schwenk, A.},
  year = {2020},
  pages = {135814},
}

@article{jiang2020,
  title = {Accurate bulk properties of nuclei from $A=2$ to $\ensuremath{\infty}$ from potentials with $\mathrm{\ensuremath{\Delta}}$ isobars},
  author = {Jiang, W. G. and Ekstr\"om, A. and Forss\'en, C. and Hagen, G. and Jansen, G. R. and Papenbrock, T.},
  journal = {Phys. Rev. C},
  volume = {102},
  issue = {5},
  pages = {054301},
  numpages = {8},
  year = {2020},
  month = {Nov},
  publisher = {American Physical Society},
  doi = {10.1103/PhysRevC.102.054301},
  url = {https://link.aps.org/doi/10.1103/PhysRevC.102.054301}
}

@article{miyagi2024,
  title = {Impact of Two-Body Currents on Magnetic Dipole Moments of Nuclei},
  author = {Miyagi, T. and Cao, X. and Seutin, R. and Bacca, S. and Ruiz, R. F. Garcia and Hebeler, K. and Holt, J. D. and Schwenk, A.},
  journal = {Phys. Rev. Lett.},
  volume = {132},
  issue = {23},
  pages = {232503},
  numpages = {6},
  year = {2024},
  month = {Jun},
  publisher = {American Physical Society},
  doi = {10.1103/PhysRevLett.132.232503},
  url = {https://link.aps.org/doi/10.1103/PhysRevLett.132.232503}
}

@article{Vernon:2010,
author = "Vernon, Ian and Goldstein, Michael and Bower, Richard G.",
doi = "10.1214/10-BA524",
fjournal = "Bayesian Analysis",
journal = "Bayesian Anal.",
month = "12",
number = "4",
pages = "619--669",
publisher = "International Society for Bayesian Analysis",
title = "Galaxy formation: a Bayesian uncertainty analysis",
volume = "5",
year = "2010"
}

@article{Vernon:2014,
author = "Vernon, Ian and Goldstein, Michael and Bower, Richard",
doi = "10.1214/12-STS412",
fjournal = "Statistical Science",
journal = "Statist. Sci.",
month = "02",
number = "1",
pages = "81--90",
publisher = "The Institute of Mathematical Statistics",
title = "Galaxy Formation: Bayesian History Matching for the Observable Universe",
volume = "29",
year = "2014"
}

@article{Hammer:2019poc,
  author  = {Hammer, H.-W. and K{\"o}nig, S. and van Kolck, U.},
  title   = {Nuclear effective field theory: Status and perspectives},
  journal = {Rev. Mod. Phys.},
  volume  = {92},
  pages   = {025004},
  year    = {2020},
  doi     = {10.1103/RevModPhys.92.025004}
}

@article{Machleidt:2011zz,
  author  = {Machleidt, R. and Entem, D. R.},
  title   = {Chiral effective field theory and nuclear forces},
  journal = {Phys. Rep.},
  volume  = {503},
  pages   = {1--75},
  year    = {2011},
  doi     = {10.1016/j.physrep.2011.02.001}
}

@article{Hergert:2020bxy,
  author  = {Hergert, H.},
  title   = {A guided tour of \textit{ab initio} nuclear many-body theory},
  journal = {Front. Phys.},
  volume  = {8},
  pages   = {379},
  year    = {2020},
  doi     = {10.3389/fphy.2020.00379}
}

@article{Owen,
  author  = {Owen, Art B.},
  title   = {Sobol' indices and Shapley value},
  journal = {SIAM/ASA J. Uncertain. Quantif.},
  volume  = {2},
  pages   = {245--251},
  year    = {2014},
  doi     = {10.1137/130936233}
}

@article{Krebs:2007rh,
  author  = {Krebs, Hermann and Epelbaum, Evgeny and Mei{\ss}ner, Ulf-G.},
  title   = {Nuclear forces with {$\Delta$}-excitations up to next-to-next-to-leading order. {I}. {P}eripheral nucleon-nucleon waves},
  journal = {Eur. Phys. J. A},
  volume  = {32},
  pages   = {127--137},
  year    = {2007},
  doi     = {10.1140/epja/i2007-10372-y}
}

@article{belley2025globalframeworkemulationnuclear,
  author  = {Belley, Antoine and Munoz, Jose M. and Garcia Ruiz, Ronald F.},
  title   = {Global framework for emulation of nuclear calculations},
  journal = {Phys. Rev. Lett.},
  volume  = {136},
  pages   = {082501},
  year    = {2026},
  doi     = {10.1103/mvc3-qdtc}
}

@article{Cook:2024toj,
  author  = {Cook, Patrick and Jammooa, Danny and Hjorth-Jensen, Morten and Lee, Daniel D. and others},
  title   = {Parametric matrix models},
  journal = {Nat. Commun.},
  volume  = {16},
  pages   = {5929},
  year    = {2025},
  doi     = {10.1038/s41467-025-61362-4}
}

@article{belley2024ab,
  author  = {Belley, Antoine and Yao, J. M. and Bally, Benjamin and Pitcher, Jack and others},
  title   = {Ab initio uncertainty quantification of neutrinoless double-beta decay in $^{76}$Ge},
  journal = {Phys. Rev. Lett.},
  volume  = {132},
  pages   = {182502},
  year    = {2024},
  doi     = {10.1103/PhysRevLett.132.182502}
}

@article{hu2022ab,
  author  = {Hu, Baishan and Jiang, Weiguang and Miyagi, Takayuki and Sun, Zhonghao and others},
  title   = {Ab initio predictions link the neutron skin of $^{208}$Pb to nuclear forces},
  journal = {Nat. Phys.},
  volume  = {18},
  pages   = {1196--1200},
  year    = {2022},
  doi     = {10.1038/s41567-022-01715-8}
}

@article{ekstrom2019global,
  author  = {Ekstr{\"o}m, Andreas and Hagen, Gaute},
  title   = {Global sensitivity analysis of bulk properties of an atomic nucleus},
  journal = {Phys. Rev. Lett.},
  volume  = {123},
  pages   = {252501},
  year    = {2019},
  doi     = {10.1103/PhysRevLett.123.252501}
}

@article{jiang2024emulating,
  author  = {Jiang, W. G. and Forss{\'e}n, Christian and Dj{\"a}rv, Tor and Hagen, G.},
  title   = {Emulating ab initio computations of infinite nucleonic matter},
  journal = {Phys. Rev. C},
  volume  = {109},
  pages   = {064314},
  year    = {2024},
  doi     = {10.1103/PhysRevC.109.064314}
}

@article{RevModPhys.81.1773,
  author  = {Epelbaum, E. and Hammer, H.-W. and Mei{\ss}ner, Ulf-G.},
  title   = {Modern theory of nuclear forces},
  journal = {Rev. Mod. Phys.},
  volume  = {81},
  pages   = {1773--1825},
  year    = {2009},
  doi     = {10.1103/RevModPhys.81.1773}
}

@article{garcia2015ground,
  author  = {Garcia Ruiz, R. F. and Bissell, M. L. and Blaum, K. and Fr{\"o}mmgen, N. and others},
  title   = {Ground-state electromagnetic moments of calcium isotopes},
  journal = {Phys. Rev. C},
  volume  = {91},
  pages   = {041304},
  year    = {2015},
  doi     = {10.1103/PhysRevC.91.041304}
}

@article{RevModPhys.93.025010,
  author  = {Tiesinga, Eite and Mohr, Peter J. and Newell, David B. and Taylor, Barry N.},
  title   = {{CODATA} recommended values of the fundamental physical constants: 2018},
  journal = {Rev. Mod. Phys.},
  volume  = {93},
  pages   = {025010},
  year    = {2021},
  doi     = {10.1103/RevModPhys.93.025010}
}

@inproceedings{ke2017lightgbm,
  author    = {Ke, Guolin and Meng, Qi and Finley, Thomas and Wang, Taifeng and others},
  title     = {{LightGBM}: A highly efficient gradient boosting decision tree},
  booktitle = {Adv. Neural Inf. Process. Syst.},
  volume    = {30},
  year      = {2017}
}

@inproceedings{akiba2019optuna,
  author    = {Akiba, Takuya and Sano, Shotaro and Yanase, Toshihiko and Ohta, Takeru and others},
  title     = {Optuna: A next-generation hyperparameter optimization framework},
  booktitle = {Proc. 25th ACM SIGKDD},
  pages     = {2623--2631},
  year      = {2019}
}

@misc{stroberg2018imsrg++,
  author       = {Stroberg, S. R.},
  title        = {imsrg++},
  howpublished = {\url{https://github.com/ragnarstroberg/imsrg}},
  year         = {2018}
}

@article{shimizu2019thick,
  author  = {Shimizu, Noritaka and Mizusaki, Takahiro and Utsuno, Yutaka and Tsunoda, Yusuke},
  title   = {Thick-restart block {L}anczos method for large-scale shell-model calculations},
  journal = {Comput. Phys. Commun.},
  volume  = {244},
  pages   = {372--384},
  year    = {2019},
  doi     = {10.1016/j.cpc.2019.06.011}
}

@article{acharya2024magnetic,
  author  = {Acharya, Bijaya and Hu, B. S. and Bacca, S. and Hagen, G. and others},
  title   = {Magnetic dipole transition in $^{48}$Ca},
  journal = {Phys. Rev. Lett.},
  volume  = {132},
  pages   = {232504},
  year    = {2024},
  doi     = {10.1103/PhysRevLett.132.232504}
}

@article{PhysRevC.83.031301,
  author  = {Hebeler, K. and Bogner, S. K. and Furnstahl, R. J. and Nogga, A. and others},
  title   = {Improved nuclear matter calculations from chiral low-momentum interactions},
  journal = {Phys. Rev. C},
  volume  = {83},
  pages   = {031301},
  year    = {2011},
  doi     = {10.1103/PhysRevC.83.031301}
}

@article{PhysRevC.91.051301,
  author  = {Ekstr{\"o}m, A. and Jansen, G. R. and Wendt, K. A. and Hagen, G. and others},
  title   = {Accurate nuclear radii and binding energies from a chiral interaction},
  journal = {Phys. Rev. C},
  volume  = {91},
  pages   = {051301},
  year    = {2015},
  doi     = {10.1103/PhysRevC.91.051301}
}

@article{Stroberg_2019,
  author  = {Stroberg, S. Ragnar and Hergert, Heiko and Bogner, Scott K. and Holt, Jason D.},
  title   = {Nonempirical interactions for the nuclear shell model: An update},
  journal = {Annu. Rev. Nucl. Part. Sci.},
  volume  = {69},
  pages   = {307--362},
  year    = {2019},
  doi     = {10.1146/annurev-nucl-101917-021120}
}

@inproceedings{lundberg2017unified,
  author    = {Lundberg, Scott M. and Lee, Su-In},
  title     = {A unified approach to interpreting model predictions},
  booktitle = {Adv. Neural Inf. Process. Syst.},
  volume    = {30},
  year      = {2017}
}

@misc{FRAME_code,
  author       = {Munoz, Jose M. and Belley, Antoine},
  title        = {{FRAME}: {F}idelity-{R}esolved {A}ffine {M}atrix {E}mulator},
  year         = {2026},
  howpublished = {\url{https://github.com/munozariasjm/FRAME-public}}
}

@article{Sanderson2016,
  doi = {10.21105/joss.00026},
  year = {2016},
  publisher = {The Open Journal},
  volume = {1},
  number = {2},
  pages = {26},
  author = {Conrad Sanderson and Ryan Curtin},
  title = {Armadillo: a template-based C++ library for linear algebra},
  journal = {Journal of Open Source Software},
  url = {https://doi.org/10.21105/joss.00026},
}

@article{Stro17ENO,
      author         = "Stroberg, S. R. and Calci, A. and Hergert, H. and Holt,
                        J. D. and Bogner, S. K. and Roth, R. and Schwenk, A.",
      title          = "{Nucleus-dependent valence-space approach to nuclear
                        structure}",
      journal        = "Phys. Rev. Lett.",
      volume         = "118",
      year           = "2017",
      pages          = "032502",
      doi            = "10.1103/PhysRevLett.118.032502",
      eprint-         = "1607.03229",
      SLACcitation   = "%%CITATION = ARXIV:1607.03229;%%"
}

@article{Parz17Trans,
      author         = "Parzuchowski, N. M. and Stroberg, S. R. and Navrátil, P.
                        and Hergert, H. and Bogner, S. K.",
      title          = "{Ab initio electromagnetic observables with the in-medium
                        similarity renormalization group}",
      journal        = "Phys. Rev. C",
      volume         = "96",
      year           = "2017",
      pages          = "034324",
      doi            = "10.1103/PhysRevC.96.034324",
}

@article{carlsson2016uncertainty,
  title={Uncertainty analysis and order-by-order optimization of chiral nuclear interactions},
  author={Carlsson, BD and Ekstr{\"o}m, A and Forss{\'e}n, Christian and Str{\"o}mberg, Dag and others},
  journal={Physical Review X},
  volume={6},
  number={1},
  pages={011019},
  year={2016},
  publisher={APS}
}

@article{4tky-t2h1,
  title = {Two-body currents at finite momentum transfer and applications to $M1$ transitions},
  author = {Brase, C. and Miyagi, T. and Men\'endez, J. and Schwenk, A.},
  journal = {Phys. Rev. C},
  volume = {113},
  issue = {1},
  pages = {014317},
  numpages = {11},
  year = {2026},
  month = {Jan},
  publisher = {American Physical Society},
  doi = {10.1103/4tky-t2h1},
  url = {https://link.aps.org/doi/10.1103/4tky-t2h1}
}

@book{CastelTowner1990,
  author    = {Castel, Boris and Towner, Ian S.},
  title     = {Modern Theories of Nuclear Moments},
  publisher = {Clarendon Press},
  address   = {Oxford},
  year      = {1990}
}

@article{Neyens2003,
  author  = {Neyens, Gerda},
  title   = {Nuclear magnetic and quadrupole moments for nuclear structure
             research on exotic nuclei},
  journal = {Rep. Prog. Phys.},
  volume  = {66},
  pages   = {633--689},
  year    = {2003},
  doi     = {10.1088/0034-4885/66/4/205}
}

@article{Yang2023,
  author  = {Yang, X. F. and Wang, S. J. and Wilkins, S. G. and
             Garcia Ruiz, R. F.},
  title   = {Laser spectroscopy for the study of exotic nuclei},
  journal = {Prog. Part. Nucl. Phys.},
  volume  = {129},
  pages   = {104005},
  year    = {2023},
  doi     = {10.1016/j.ppnp.2022.104005}
}

@article{Martin2023,
  author  = {Martin, J. D. and Novario, S. J. and Lonardoni, D. and
             Carlson, J. and Gandolfi, S. and Tews, I.},
  title   = {Auxiliary field diffusion {M}onte {C}arlo calculations of
             magnetic moments of light nuclei with chiral effective field
             theory interactions},
  journal = {Phys. Rev. C},
  volume  = {108},
  pages   = {L031304},
  year    = {2023},
  doi     = {10.1103/PhysRevC.108.L031304}
}

@article{Gysbers2019,
  author  = {Gysbers, P. and Hagen, G. and Holt, J. D. and Jansen, G. R. and
             Morris, T. D. and Navr{\'a}til, P. and Papenbrock, T. and
             Quaglioni, S. and Schwenk, A. and Stroberg, S. R. and
             Wendt, K. A.},
  title   = {Discrepancy between experimental and theoretical
             $\beta$-decay rates resolved from first principles},
  journal = {Nat. Phys.},
  volume  = {15},
  pages   = {428--431},
  year    = {2019},
  doi     = {10.1038/s41567-019-0450-7}
}

@article{Sassarini_2022,
doi = {10.1088/1361-6471/ac900a},
url = {https://doi.org/10.1088/1361-6471/ac900a},
year = {2022},
month = {sep},
publisher = {IOP Publishing},
volume = {49},
number = {11},
pages = {11LT01},
author = {Sassarini, P L and Dobaczewski, J and Bonnard, J and Ruiz, R F Garcia},
title = {Nuclear DFT analysis of electromagnetic moments in odd near doubly magic nuclei},
journal = {Journal of Physics G: Nuclear and Particle Physics}
}

@misc{Wang2026,
  author        = {Wang, Teng and Elhatisari, Serdar and Feng, Xu and Lee, Dean
                   and Lu, Bing-Nan and Ma, Yuan-Zhuo},
  title         = {Ab initio lattice calculation of nuclear magnetic dipole moments with systematic error quantifications},
  year          = {2026},
  eprint        = {2607.25464},
  archivePrefix = {arXiv},
  primaryClass  = {nucl-th},
  doi           = {10.48550/arXiv.2607.25464}
}

\section*{End Matter}

\textbf{Electromagnetic operators.}
We use the standard decomposition of the magnetic dipole operator into one-body currents (1BC) and two-body currents (2BC)
\begin{equation}
\boldsymbol{\mu}=\boldsymbol{\mu}_{\mathrm{1B}}+\boldsymbol{\mu}_{\mathrm{2B}},\qquad
\mu=\langle J M\!=\!J|\,\mu_z\,|J M\!=\!J\rangle.
\end{equation}
The one-body term is given by
\begin{equation}
\mu_{\mathrm{1B},z}=\mu_N\sum_i\big(g^\ell_i\,\ell_{i,z}+g^s_i\,\sigma_{i,z}\big),\quad
\mu_N=\frac{e\hbar}{2m_p},
\end{equation}
using free-nucleon $g$-factors~\cite{RevModPhys.93.025010}. The leading 2BC are taken from $\chi$EFT at the same order as the nuclear interaction (N$^2$LO) and include both intrinsic and Sachs terms, as detailed in Refs.~\cite{miyagi2024,4tky-t2h1}.

The electric quadrupole ($Q$) moment is obtained from the reduced matrix element of the evolved quadrupole operator,
\begin{equation}
Q(J)=\sqrt{\frac{16\pi}{5}}\;\frac{\langle J || \hat{Q}^{(2)} || J \rangle}{\sqrt{2J+1}},\qquad
\hat{Q}^{(2)}=\sum_i e_i\,r_i^2\,Y_2(\hat{r}_i).
\end{equation}
Initial operators are normal-ordered with respect to the finite-nucleus reference state and evolved consistently at the IMSRG(2) level.

\textbf{Emulator training set.}
The open-source implementation of the emulator is available in Ref.~\cite{FRAME_code}. Directly propagating this extensive ensemble through many-body solvers is computationally intractable, so instead, we construct a high-fidelity training set by solving the valence-space in-medium similarity renormalization group (VS-IMSRG)~\cite{Stro17ENO,Stroberg_2019} equations for a representative subset of interactions. We use the \texttt{imsrg++}~\cite{stroberg2018imsrg++} code for the consistent evolution of the Hamiltonian and operators, followed by exact diagonalization via the shell-model code \texttt{KSHELL}~\cite{shimizu2019thick}. These calculations are performed across a hierarchy of single-particle model-space truncations ($e_{\max}\!\in\!\{4,6,8,10\}$), where single-particle states are characterized by $e = 2n+l$ for the principal quantum number $n$ and orbital angular momentum $l$.

\textbf{Posterior-predictive marginalization.}
\begin{gather*}
 p(Y \mid \mathcal{D}) = \int g(Z, N, f, \boldsymbol{\alpha})\, p(\boldsymbol{\alpha} \mid \mathcal{D})\, d\boldsymbol{\alpha} \;\\\approx\; \sum_{i} w_i \, g(Z, N, f, \boldsymbol{\alpha}_i),
    \label{eq:posterior_predictive}
\end{gather*}
where $\mathcal{D}$ denotes the calibration data, $g$ is the FRAME emulator, and $w_i$ are the normalized posterior weights derived from the likelihood being used.

\textbf{Sensitivity methodology.}
By contrast, our posterior-integrated approach restricts the sensitivity analysis to the history-matching informed manifold. This ensures that we attribute importance only to parameter variations that maintain fidelity to few-body data, providing a physically faithful picture of how the constrained interaction drives the spread of many-body observables. Algorithmic and statistical details of the methods are given below.

\textbf{Posterior importance estimation.}
For a fixed nucleus $(Z,N)$, fidelity $f$, and observable $Y\in\{E_b,R_{\mathrm{ch}},\mu,Q\}$, the emulator $g:(Z,N,f,\boldsymbol{\alpha})\mapsto Y$ maps a LEC vector $\boldsymbol{\alpha}\in\mathbb{R}^d$ to $Y$. The LEC ensemble $\{\boldsymbol{\alpha}_i\}_{i=1}^M$ is obtained from history matching and reweighted by a likelihood over calibrating observables. We denote the normalized posterior weights by $w_i\ge 0$ with $\sum_i w_i=1$ and define the effective sample size
\begin{equation}
  \mathrm{ESS}=\frac{1}{\sum_{i=1}^M w_i^2}.
\end{equation}

\textbf{Posterior over LECs.}
Starting from a non-implausible set produced by history matching, we apply a discrete Bayes/SIR update using a likelihood built from calibrating observables. The resulting weights are normalized to $w_i$ and used for all posterior aggregations below.

\textbf{Surrogate for local attribution.}
Shapley attributions require a model that is cheap to perturb. For each $(Z,N,f,Y)$ we train a gradient-boosted decision tree surrogate $\widehat{g}$ on a posterior-resampled LEC cloud:
\begin{enumerate}
\item Partition $\{(\boldsymbol{\alpha}_i,w_i)\}$ into training and test sets.
\item Produce an augmented training set by weighted resampling with small per-LEC Gaussian jitter. The jitter scale is set by a robust posterior spread (interquartile range or, if zero, posterior standard deviation).
\item Recompute labels $\widehat{Y}=g(\boldsymbol{\alpha};Z,N,f)$ with the emulator on the augmented cloud.
\item Fit $\widehat{g}$ on $(\boldsymbol{\alpha},\widehat{Y})$ with fixed hyperparameters across nuclei.
\item Audit the fit with posterior-weighted $R^2_w$ and Root Mean Squared Error (RMSE) ${\rm RMSE}_w$ on the held-out posterior split; we require $R^2_w\ge 0.9$ before reporting importances.
\end{enumerate}

For this implementation, we use LightGBM~\cite{ke2017lightgbm} gradient-boosted trees with fixed hyperparameters across nuclei, trained on a posterior-resampled cloud (up to $10^5$ rows). For each partition $\{(\boldsymbol{\alpha}_i,w_i)\}$, the fitting parameters are hyperoptimized with Optuna~\cite{akiba2019optuna} to avoid overfitting of the regressors. All importance values reported in this work are averaged over ten independent random initializations of the emulator training and of the attribution pipeline. The dispersion across initializations is small compared to the isotope-to-isotope variation discussed in the main text, and all conclusions are qualitatively unchanged across initializations.

\textbf{SHAP definitions and computation.}
We use two complementary Shapley regimes for $\widehat{g}$.

\textbf{Interventional SHAP (main effects).}
Let $\mathcal{B}$ be a background matrix drawn from the posterior over $\boldsymbol{\alpha}$. Interventional SHAP independently perturbs feature $j$ relative to $\mathcal{B}$. For evaluation set $X$ with per-row weights $w(x)$ (posterior-proportional), per-feature attributions $\phi^{\rm int}_j(x)$ are aggregated as
\begin{equation}
  \bar{\phi}^{\rm int}_j=\sum_{x\in X} w(x)\,\phi^{\rm int}_j(x).
\end{equation}
\begin{equation}
    S^{\rm main}_j=\frac{\sum_{x\in X} w(x)\,|\phi^{\rm int}_j(x)|}{\sum_{k=1}^d \sum_{x\in X} w(x)\,|\phi^{\rm int}_k(x)|}
\end{equation}
$S^{\rm main}_j$ is reported as the main-effect share.

\textbf{Path-dependent SHAP (main + interactions).}
Using tree-path perturbations, we obtain per-feature $\phi_j(x)$ and pairwise interaction tensors $\Phi_{jk}(x)$. Posterior aggregation yields the main absolute contributions
\begin{equation}
  M_j=\sum_{x\in X} w(x)\,|\Phi_{jj}(x)|,
\end{equation}
and totals
\begin{equation}
  T_j=M_j+\sum_{k\neq j}\sum_{x\in X} w(x)\,|\Phi_{jk}(x)|.
\end{equation}
We report $S^{\rm tot}_j=T_j\big/\sum_{\ell=1}^d T_\ell$ and rank interaction pairs by $\sum_x w(x)\,|\Phi_{jk}(x)|$.

\textbf{Uncertainty budget.} The total variance combines experimental and emulator uncertainties and two dominant sources of systematic uncertainty due to the truncation of $\chi$EFT and the many-body method,
\begin{equation}
\sigma_{k,\text{tot}}^2 = \sigma_{\text{exp}}^2 + \sigma_{\text{emu}}^2 + \tau_{\chi}^2 + \sigma_{\text{MBT}}^2.
\end{equation}
We estimate the former using an observed spread in the predictions for the magnetic dipole moment in calcium isotopes~\cite{acharya2024magnetic} using the established interactions 1.8/2.0 (EM)~\cite{PhysRevC.83.031301}, $\Delta$NNLO$_{\rm{{GO}}}$(394)~\cite{jiang2020}, and NNLO$_{\mathrm{sat}}$~\cite{PhysRevC.91.051301}. For the many-body truncation uncertainty, we take half the difference between our major-shell calculation and available multi-shell results for $\Delta$NNLO$_{\rm{{GO}}}$(394). Operator uncertainties are treated as subleading (see below).

\textbf{Likelihood and error budget.}
The likelihood function assumes independent Gaussian errors for the calibration isotopes $N \in \{19, 27, 29, 31\}$:
\begin{equation}
    \ln \mathcal{L} = -\frac{1}{2} \sum_{k} \left[ \frac{( \mu_k^{\rm exp} - \mu_k^{\rm emu}(\boldsymbol{\alpha}) )^2}{\sigma_{k, \rm tot}^2} + \ln(2\pi \sigma_{k, \rm tot}^2) \right].
\end{equation}
For this work, we adopt $\tau_{\chi} \approx 0.04 \mu_N$ (EFT truncation error) and a nuclide-dependent many-body truncation error $\sigma_{\rm MBT}$ derived from the spread between single-shell and multi-shell VS-IMSRG calculations. The emulator uncertainty $\tau_{\rm emu}$ is taken as the root-mean-square residual of the emulator against the VS-IMSRG training set for that nuclide at $e_{\max}=10$.

\textbf{SIR implementation.}
The posterior distribution over the LECs, $p(\boldsymbol{\alpha} | D_{\rm new})$, is approximated using a Sampling Importance Resampling (SIR) scheme applied to the history-matched ensemble. The updated weights $w_i^{\rm post}$ for each sample $i$ are computed as:
\begin{equation}
    w_i^{\rm post} \propto \left( w_i^{\rm prior} \right)^{1/T} \times \mathcal{L}(\mathbf{y}_{\rm Ca} | \boldsymbol{\alpha}_i),
\end{equation}
where $\mathcal{L}$ is the likelihood function for the calcium magnetic moments and $T$ is a temperature hyperparameter applied to the prior weights. The base prior weights $w_i^{\rm prior}$ are derived from the original history matching using the $\mathcal{L}_{(2-4)+^{16}\text{O}}$ likelihood~\cite{jiang2020,jiang2024emulating}.

To ensure the analysis is not dominated by the specific choice of the prior tempering (which controls the diffusiveness of the initial history-matched cloud), we perform a sensitivity analysis by varying $T$. The baseline results presented in the main text use $T=5.0$, selected to maximize the effective sample size (ESS) while retaining the structural information of the prior. We validate the robustness of our conclusions by varying $T \in [3.0, 7.0]$, over which the shrinkage pattern and all reported trends remain stable.

\end{document}